\documentclass{INTERSPEECH2023}

\usepackage{multirow}
\usepackage{mathtools}
\usepackage{caption,subcaption}

\interspeechcameraready

\title{Robust Self Supervised Speech Embeddings for Child-Adult Classification in Interactions involving Children with Autism}
\vspace{-10ex}
\name{Rimita Lahiri$^1$, Tiantian Feng$^1$, Rajat Hebbar$^1$, Catherine Lord$^2$, So Hyun Kim$^3$, Shrikanth Narayanan$^1$}
\vspace{-10ex}
\vspace{-10ex}
\address{
  $^1$Signal Analysis and Interpretation Laboratory, University of Southern California, USA\\
  $^2$Semel Institute of Neuroscience and Human Behavior, University of California, USA\\
  $^3$School of Psychology, Korea University, Korea}
\email{rlahiri@usc.edu}

\begin{document}

\maketitle
 
\begin{abstract}
We address the problem of detecting who spoke when in child-inclusive spoken interactions i.e., automatic child-adult speaker classification. 
Interactions involving children are richly heterogeneous due to developmental differences. The presence of neurodiversity e.g., due to Autism, contributes additional variability. We investigate the impact of additional pre-training with more unlabelled child speech on the child-adult classification performance. 
We pre-train our model with child-inclusive interactions, following two recent self-supervision algorithms, Wav2vec 2.0 and WavLM, with a contrastive loss objective. We report  $9 - 13\%$  relative improvement over the state-of-the-art baseline with regards to classification F1 scores on two clinical interaction datasets involving children with Autism. 
We also analyze the impact of pre-training under different conditions by evaluating our model on interactions involving different subgroups of children based on various demographic factors.
 
\end{abstract}

\vspace{-2ex}
\section{Introduction}
\label{section:intro}
\vspace{-1ex}

\textit{Autism Spectrum Disorder}~(ASD) is a neuro-developmental disorder, characterized by deficits in social and communicative abilities along with restrictive repetitive behavior~\cite{volden1991neologisms,huemer2010comprehensive}. Individuals with ASD tend to show symptoms of anomalies in language, non-verbal comprehension, expressions and vocal prosody patterns~\cite{kim2014language, sorensen2019cross}. In the United States, the prevalence of ASD in children has steadily increased from 1 in 150~\cite{centers2006mental} in 2002 to 1 in 44 in 2022. It is critical to develop early ASD diagnosis to create timely interventions. One of the most common observation tools supporting ASD diagnostic and intervention efforts includes clinically-administered semi-structured dyadic interactions between the child and a trained clinician~\cite{lord2000autism,grzadzinski2016measuring}. Computational analysis of such interactions provides evidence-driven opportunities for the support of behavioral stratification as well as diagnosis and personalized treatment. 


\begin{figure}
    \centering
    \includegraphics[width=0.47\textwidth]{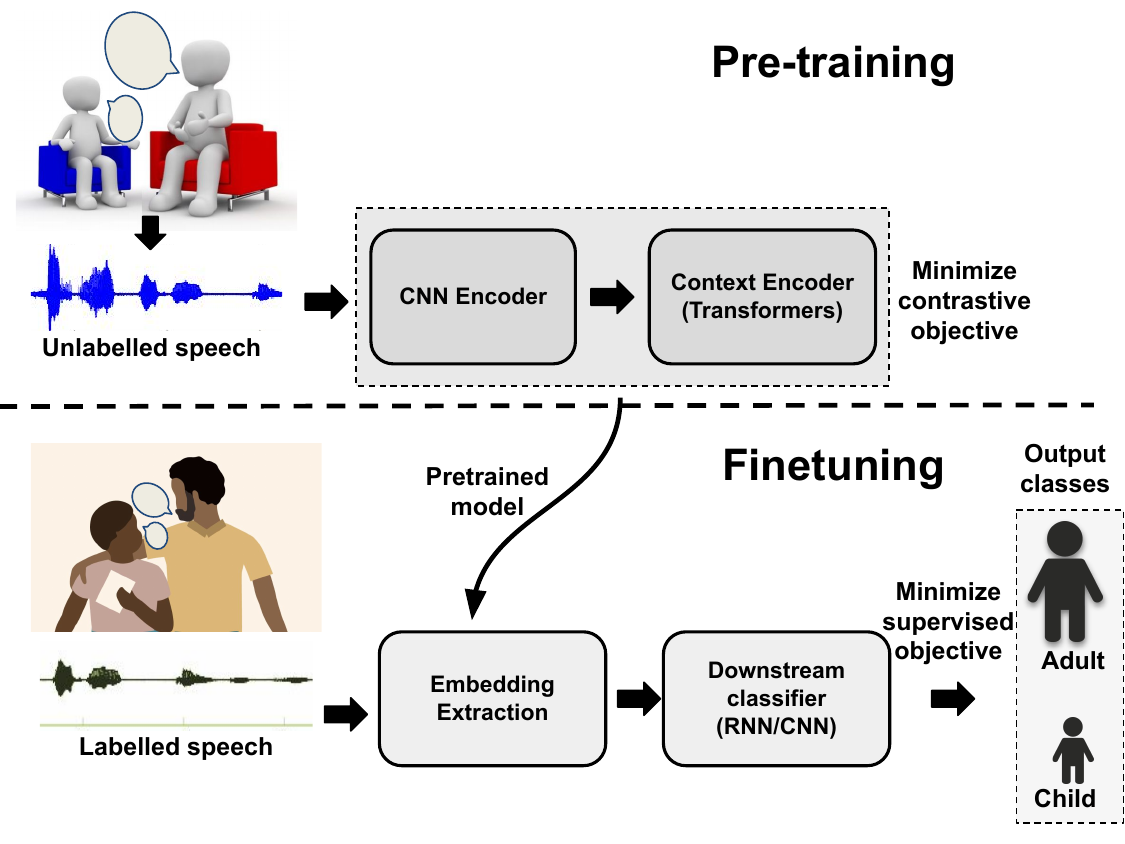}
    \vspace{-5.5ex}
    \captionsetup{justification=centering}
    \caption{Schematic overview of the proposed two-step recipe for child-adult classification}
    \label{fig:recipe}
\vspace{-5.5ex}
\end{figure}

\par However, with regards to behavioral feature extraction and analysis for these dyadic interactions, prior works have primarily relied on human-annotated data segmentation by speaker labels, which is expensive and time-consuming to obtain, especially for large corpora. Computational modeling of naturalistic conversations has gained a lot of attention in the past few decades because of its potential in rich human behavioral phenotyping. Hence, it is desirable to conduct automatic analysis of these interactions using signal processing and machine learning. Specifically, one fundamental module for supporting automated processing of child-adult interactions is the task of child-adult speech classification i.e., distinguishing the speech regions of the child from those of an interacting adult. Analysis of child speech is more challenging than adult speech because of the wide variability and idiosyncrasies associated with child \cite{lee1999acoustics,bhardwaj2022automatic,GURUNATHSHIVAKUMAR2022101289}. An additional layer of complexity arises while analyzing speech for the clinical domain, as different clinical conditions may lead to unique patterns in language and speech, making it challenging for current computational approaches to capture.



\par Training a robust child-adult classifier is challenging for two main factors: the scarcity of reliably labeled datasets containing child speech and the larger within-class variability due to the changes in child speech based on demographic factors like age, gender, and developmental status including any clinical symptom severity~\cite{lahiri2020learning}. Most recent works addressing the problem of speaker diarization have primarily targeted fine-tuning the pre-trained models by optimizing a supervised objective. 
So far, \textit{Self-Supervised Learning}~(SSL) algorithms are largely under-explored for leveraging unlabelled child speech for developing speaker discriminative embeddings, especially in real-world settings such as clinical diagnostic and monitoring sessions. Specifically, there is a limited understanding of how the performance of these models varies across children with different demographics, including age and gender.

\noindent \textbf{Contributions of this paper:} We address the above questions by evaluating the impact of including more child speech, during pre-training on the child-adult speaker classification. We choose Wav2vec 2.0~(W2V2)-base and WavLM-base+ as the backbone models. The detailed contributions of this work are summarized as:
\begin{itemize}[leftmargin=*]

\item Our work represents one of the first attempts to \textbf{leverage unlabelled child speech in pre-training} for developing speaker discriminative embeddings, especially due to the vast inherent heterogeneity in the data arising from developmental differences.
\item We experimentally substantiate the effectiveness of our method for downstream child-adult speaker classification tasks using W2V2 and WavLM and \textbf{report over  13\% and 9\% relative improvement over the base models in terms of F1 scores} in two datasets, respectively.
\item We also illustrate and analyze the performance of the proposed method among different subgroups of children based on demographic factors. 

\end{itemize}


\vspace{-3ex}
\section{Background}
\vspace{-1.5ex}
\subsection{Self-supervision in speech}
\vspace{-1.5ex}

The need for building speech processing frameworks in low/limited resource scenarios has spurred significant efforts on unsupervised, semi-supervised and weakly supervised learning strategies to reduce reliance on labeled datasets. The success of SSL~\cite{devlin2018bert} in natural language processing, notably due to its generalizability and transferability, has also inspired its adoption within the speech domain. Early studies explored SSL in speech with generative loss~\cite{liu2021tera,ling2020decoar}, while more recent ones have focused on discriminative loss~\cite{baevski2020wav2vec,schneider2019wav2vec} and multi-task learning objectives~\cite{pascual2019learning,ravanelli2020multi}. The current approach in this realm follows a two-step process: first pre-train a model in a self-supervised manner on large amounts of unlabeled data to encode general-purpose knowledge, and next specialize the model on various downstream tasks through fine-tuning. Past studies have reported the efficacy of SSL algorithms by leveraging the pre-trained embeddings on downstream tasks including ASR~\cite{baevski2020wav2vec}, speaker verification~\cite{fan2020exploring}, speaker identification~\cite{chen2022wavlm}, phoneme classification~\cite{chung2019unsupervised}, emotion recognition~\cite{wang2021fine}, spoken language understanding~\cite{wang2021fine}, and TTS~\cite{alvarez2019problem}. 



\vspace{-1.5ex}
\subsection{Child-adult classification in the ASD domain}
\vspace{-1ex}

Child-adult classification is among the more difficult tasks within speaker diarization, due to the challenges related to "in the wild" child speech in naturalistic conversational settings including short speaker turns, varied noise sources and a larger fraction of overlapping speech. Early diarization solutions involving child speech
used traditional feature representations~(MFCCs, PLPs)~\cite{najafian2016speaker, cristia2018talker}. In \cite{zhou2016speaker}, the authors introduced several methods for processing audio collected from children with autism using a wearable device. Later, deep speech representations, i-vectors~\cite{zhou2016speaker} and x-vectors~\cite{xie2019multi} were studied for this task. A variety of challenges, both from signal processing and limited data availability, have been identified and addressed. In ~\cite{lahiri2020learning}, the authors have proposed an adversarial training strategy to address the large within- and across-age and gender variability due to developmental changes in children. Alternatively, in ~\cite{koluguri2020meta}, pre-trained x-vectors were fine-tuned for child/adult speaker diarization using a meta-learning paradigm, namely prototypical networks. Moreover, the role of the amount of child speech in building deep neural speaker representations was studied in ~\cite{krishnamachari2021developing} and their experimental results confirm that including more child data indeed enhances the task performance in a supervised setup.

\vspace{-3.5ex}
\section{Datasets}
\label{sec:data}
\vspace{-1ex}
Our child-inclusive data come from interactions in a clinical setting, specifically obtained during the administration of two clinical protocols related to developmental disorders. The first protocol is the gold standard \textit{Autism Diagnostic Observation Schedule}~(ADOS)~\cite{lord2000autism}, used for diagnostic purposes. The second protocol is a recently proposed outcome-measure focused instrument \textit{Brief Observation of Social Communication Change (BOSCC)}~\cite{grzadzinski2016measuring} for tracking changes in social and communicative skills during the course of treatment. A typical ADOS session lasts $40-60$ minutes and contains multiple (usually $10-15$) semi-structured activities for addressing specific symptoms related to ASD. Usually these interactions aim to elicit spontaneous responses from children under different circumstances to obtain a diagnostic score for classifying children with and without ASD. A BOSCC session is usually 12 minutes long, consisting of two $2 min$ conversational talk sessions and two $4 min$ play sessions where the child plays with a toy. 

\par In our pre-training experiments, we use a dataset consisting of 369 recordings of unlabelled BOSCC sessions comprising approximately~$100K$ utterances. For the fine-tuning experiments, we use two different corpora, ADOSMod3 and Simons. The ADOSMod3 corpus was collected across 2 clinical sites.
These data are from administrations of the ADOS Module-3 designed for verbally fluent children, with a focus on the Social Difficulties and Annoyance and Emotional sub-tasks for this work. The data consist of total 346 sessions collected from $165$ children ($86$ ASD, $79$ Non-ASD).
The Simons corpus used in our study consists of a combination of clinically administered ADOS~($n=6$) and BOSCC~($n=33$) sessions collected across 4 sites and these sessions were labeled by trained annotators to extract speaker timestamps. The details of datasets are reported in Table~\ref{tab:dataset}.

\begin{table}[t!]
\tiny
\begin{center}
\captionsetup{justification=centering}
\caption{Session-level statistics of child-adult corpora.}
\vspace{-3ex}
\label{tab:dataset}
\resizebox{0.48\textwidth}{!}{
\begin{tabular}{c c c} 
\hline \hline
\multirow{2}{*}{\textbf{Dataset}} & \textbf{Duration} & \textbf{Child-speaking fraction} \\
& ($mean \pm std$) & ($mean \pm std$) \\
\hline \hline
Pre-training & $14.05 \pm 2.08$ & n.a \\
ADOSMod3 & $3.23 \pm 1.61$ & $0.46 \pm 0.18$ \\
Simons & $19.05 \pm 12.86$ & $0.40 \pm 0.08$ \\
\hline \hline
\end{tabular}}
\end{center}
\vspace{-10ex}
\end{table}

\vspace{-2ex}
\section{System Description}
\label{section:system}

\vspace{-0.5ex}
\subsection{Pre-training}
\label{section:pretraining}
\vspace{-1.5ex}

Our research aims to adapt the existing self-supervised approaches to the child-adult interaction domain through contrastive learning. Similar to \cite{sachidananda2022calm}, our contrastive learning framework is based on the assumption that neighboring segments from audio samples are highly likely to contain identical information. For instance, it is probable that adjacent audio frames are produced by the same speaker and are expected to contain similar semantic meaning, linguistic content, as well as acoustic characteristics. To elaborate, we define the dataset of audio samples as $N$, where each audio sample is denoted as $x_{i}$. The corresponding neighboring audio segment is represented as $x'_{i}$, and is defined as any audio sample that has a time shift of half a second or less from the original sample $x_{i}$.

As outlined in the previous section, transformer-based models first transform the input speech sample $x$ to intermediate features ${z}$ using the feature encoder $f(\cdot)$ on the basis of CNNs. Subsequently, the transformer encoder $g(\cdot)$ maps the features ${z}$ to contextualized representations ${c}$. Consequently, we can create similar pairs of contextualized representations $c_{i}$ and $c'_{i}$ from the neighboring audio segments $x_{i}$ and $x'_{i}$, with the remaining pairs being considered as negative pairs:
\vspace{-1ex}
\begin{equation}
    \text{Positive Pairs}: c_{i} \approx c'_{i}
\end{equation}
\vspace{-4ex}
\begin{equation}
    \text{Negative Pairs}: c_{i} \neq c_{k} \text{ , } c_{i} \neq c'_{k}, \text{where } i \neq k
\end{equation}

Motivated by SimCLR~\cite{chen2020simple}, we apply the NTXent contrastive loss \cite{sohn2016improved} as the pretraining objective with the adult-child conversational corpora. Given the temperature value $\tau$, the loss function $L_{NTXent}$ for the positive audio pairs $x_{i}$ and $x'_{i}$ within a batch of $B$ input audio is:

\vspace{-5ex}
\begin{equation}
    \scriptsize
    - log {\frac{exp(sim(c_{i}, c'_{i})/\tau)}{\sum^{B}_{\substack{k=0 \\ k\neq i}}exp(sim(z_{i},z_{k})/\tau) + \sum^{B}_{\substack{k=0 }}exp(sim(z_{i},z'_{k})/\tau)}}
\end{equation}

\vspace{-8.5ex}
\subsection{Downstream Classifier Architectures}
\vspace{-1ex}

We use two different neural network models for child-adult speaker classification based on \cite{feng2023trustser}. Both the classifiers include a self-attention based projector module, whereas one of them uses CNNs to capture speaker characteristics and the other uses \textit{Recurrent Neural Netowrks}~(RNN) to model the temporal dependencies present in the signal.

\par The RNN-based classifier consists of a stacked sequence of a \textit{Feed Forward Layer}~(FFL), a bidirectional~\textit{Long Short Term Memory}~(LSTM) layer, a self-attention based projector layer and an output layer comprised of 2 FFLs, separated by a non-linear activation. The CNN classifier architecture is comprised of a weighted feature extraction module, followed by a convolutional module having 3 1D convolutional layers, each with a dropout and a non-linear activation in between, a self-attention based projector layer and an output layer comprised of 2 FFLs, separated by a non-linear activation. For all the experiments we use ~\textit{Rectified Linear Unit~(ReLU)} as the non-linear activation and a dropout ratio of $0.3$.



\vspace{-2ex}
\section{Experimental Setup}
\vspace{-0.5ex}
\subsection{Child Adult Classification}
\vspace{-1ex}

In this study, we hypothesize leveraging unlabelled child-speech for pre-training can guide models to learn the heterogeneous child speech and interaction patterns, leading to enhanced performance of downstream child-adult speaker classification. Instead of training from scratch, we pre-train the existing W2V2 and WavLM models with additional unlabelled child speech by unfreezing and updating specific transformer layers using a contrastive loss described in Sec~\ref{section:pretraining}. We report the child-adult classification macro F1-score on two labeled child-adult interaction corpus described in section \ref{sec:data}. We report the results in Table~\ref{tab:W2V2results} and Table~\ref{tab:WavLMresults}, where the first row denotes downstream child-adult classification performance using the model solely relying on pre-trained embeddings. The subsequent rows denote downstream task performance using the models pre-trained with additional child speech, where the number indicates the number of trainable transformer layers involved in the pre-training task. 

\begin{table}[t!]
\tiny
\begin{center}
\captionsetup{justification=centering}
\caption{Number of trainable parameters for the pre-training experiments based on unfrozen transformer layers}
\vspace{-3ex}
\label{tab:parameters}
\resizebox{0.48\textwidth}{!}{
\begin{tabular}{c | c | c | c | c}
\hline \hline
\multicolumn{5}{c}{\textbf{Number of unfrozen transformer layers}}\\
\hline
1 & 2 & 3 & 4 & 5 \\
\hline 
6.2M & 13.5M & 20.8M & 27.1M & 33.8M\\
\hline \hline
\end{tabular}}
\end{center}
\vspace{-9ex}
\end{table}

\vspace{-2ex}

\subsection{ADOSMod3 Experiments on Demographics}
\label{sec:exp_demographic}

\begin{table}[t!]
\tiny
\begin{center}
\captionsetup{justification=centering}
\caption{Child-adult classification F1 score using W2V2. PT corresponds to pre-training and the following number represents the number of layers used for pre-training. }
\label{tab:W2V2results}
\vspace{-2.5ex}
\resizebox{0.48\textwidth}{!}{
 \begin{tabular}{c | c  | c | c | c } \hline \hline
\multirow{2}{*}{\textbf{Model}} & \multicolumn{2}{c|}{\textbf{ADOSMod3}} &  \multicolumn{2}{c}{\textbf{Simons}}\\
\cline{2-5}
& RNN & CNN & RNN & CNN \\
\hline \hline
W2V2 - Base  & 67.92 & 70.59 & 63.41 & 64.13 \\
W2V2 - PT1 & 69.31 & 72.41 & 65.19 & 66.28 \\
W2V2 - PT2 & 71.55 & 72.95 & 65.87 & 65.12 \\
W2V2 - PT3 & 72.23 & 74.38 & \textbf{68.81} & 65.44 \\
W2V2 - PT4 & \textbf{74.01} & \textbf{74.89} & 67.63 & \textbf{66.79} \\
W2V2 - PT5 & 72.19 & 74.05 & 65.01 & 65.39 \\
\hline \hline
\end{tabular}}
\end{center}
\vspace{-9ex}
\end{table}

\begin{table}[t!]
\tiny
\begin{center}
\captionsetup{justification=centering}
\caption{Child-adult classification F1 score using WavLM pre-training. PT corresponds to pre-training and the following number represents the number of layers used for pre-training.}
\label{tab:WavLMresults}
\vspace{-2.5ex}
\resizebox{0.48\textwidth}{!}{
\begin{tabular}{c | c  | c | c | c } \hline \hline
\multirow{2}{*}{\textbf{Model}} & \multicolumn{2}{c|}{\textbf{ADOSMod3}} &  \multicolumn{2}{c}{\textbf{Simons}}\\
\cline{2-5}
& RNN & CNN & RNN & CNN \\
\hline \hline
WavLM-Base & 72.73 & 73.09 & 71.78 & 70.25 \\
WavLM - PT1 & 74.29 & 74.93 & 72.64 & 71.11 \\
WavLM - PT2 & \textbf{76.66} & 75.81 & \textbf{72.88} & \textbf{72.74} \\
WavLM - PT3 & 75.95 & \textbf{76.37} & 72.31 & 71.09 \\
WavLM - PT4 & 75.18 & 75.92 & 72.01 & 71.59 \\
WavLM - PT5 & 75.48 & 73.17 & 71.47 & 70.17 \\
\hline \hline
\end{tabular}}
\end{center}
\vspace{-12ex}
\end{table}

\vspace{-1.5ex}
\begin{figure*}[!t]
\captionsetup[subfigure]{justification=centering}
    \begin{subfigure}[b]{0.24\textwidth}
        \includegraphics[width=\textwidth]{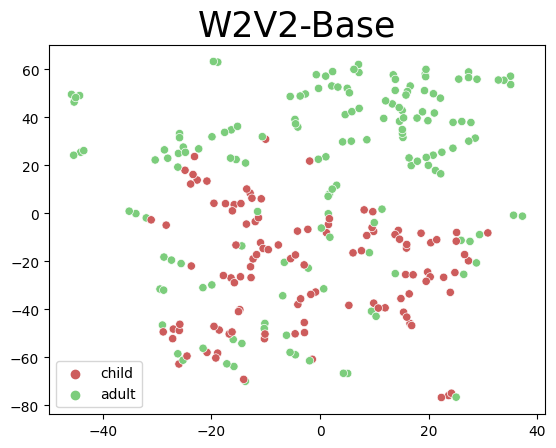}
        \vspace{-1.5ex}
        \caption{Session A:~W2V2-Base}\label{fig1.a}
    \end{subfigure}
    \begin{subfigure}[b]{0.24\textwidth}
        \includegraphics[width=\textwidth]{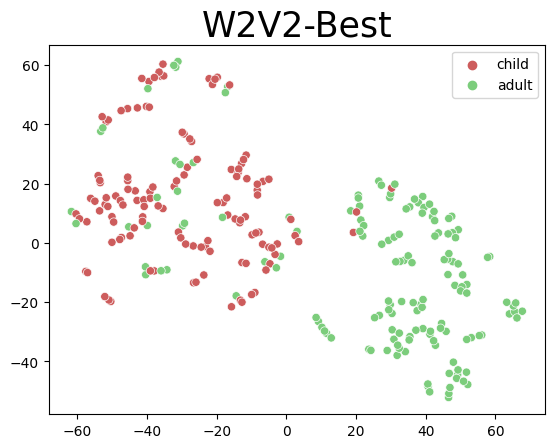}
        \vspace{-1.5ex}
        \caption{Session A:~W2V2-PT4}\label{fig1.b}
    \end{subfigure}
    \begin{subfigure}[b]{0.24\textwidth}
        \includegraphics[width=\textwidth]{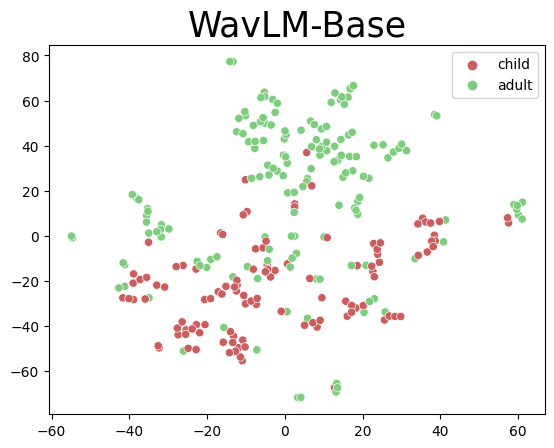}
        \vspace{-1.5ex}
        \caption{Session B:~WavLM-Base}\label{fig1.c}
    \end{subfigure}
    \begin{subfigure}[b]{0.24\textwidth}
        \includegraphics[width=\textwidth]{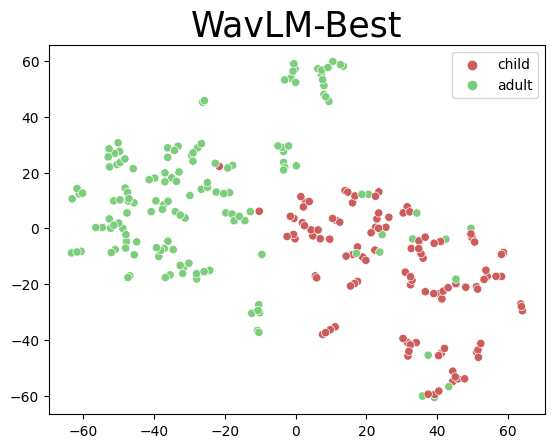}
        \vspace{-1.5ex}
        \caption{Session B:~WavLM-PT2}\label{fig1.d}
    \end{subfigure}
    \vspace{-1.5ex}
    \caption{t-SNE plots of the most discriminative 2 components of the embedding space corresponding to the classes}
    \label{fig:tsne}
\vspace{-3.5ex}
\end{figure*}

Our study also investigates the model performance across age-groups in the ADOSMod3 corpus. Prior works~\cite{lahiri2020learning,lee1999acoustics} have reported age as an important variability factor impacting speech characteristics. Based on this hypothesis, we conduct an experiment by partitioning the ADOSMod3 corpus~($3-13yrs$) into three different age-groups~(Age-group 1: 43-90 months, Age-group 2: 91-118 months, Age-group 3: 119-158 months), such that each group contains equal number of sessions. For each of these groups, we report the child-adult classification F1 score using the pre-trained base models of W2V2 and WavLM and also the best-performing pre-trained models of those two categories.

\par Not only age, analyses of developmental changes in speech have revealed sex ("gender") differences in speech characteristics, especially post puberty~\cite{lee1999acoustics}. In this work, we also report gender-based child-adult speaker classification performance on ADOSMod3 dataset, with recordings from $244$ male and $84$ female individuals. Similar to age-focused experiments, the dataset is partitioned into male and female subsets, and comparisons are drawn between the base model and the best-performing pre-trained models for both W2V2 and WavLM.

\vspace{-1ex}
\subsection{Experimental details}
\vspace{-1ex}

For both the pre-training and fine-tuning experiments, $Adam$ optimizer is used with a batch size of $32$ samples and temperature is set to $0.1$. The number of tunable parameters for the pre-training experiments is reported in Table~\ref{tab:parameters}. The initial learning rate is set to 1e-5 and the models are trained for $30$ epochs with an early stopping callback on validation loss, patience being $5$ epochs. For the downstream child-adult classification task, the model is trained to minimize the binary cross-entropy loss for a maximum of $50$ epochs, while the initial learning rate for this experiment is 2e-4 with a weight decay of 1e-4. For both the datasets, we use 70\% for training, 15\% for validation and 15\% for testing. We use the model checkpoints from HuggingFace~\cite{wolf2020transformers}. We pre-train the models using a single NVIDIA GeForce GPU 1080 Ti and each experiment took less than two days.

\vspace{-4ex}
\section{Results and Discussion}
\label{section:results}

\begin{figure}
    \centering
    \captionsetup{justification=centering}
    \includegraphics[width=0.45\textwidth]{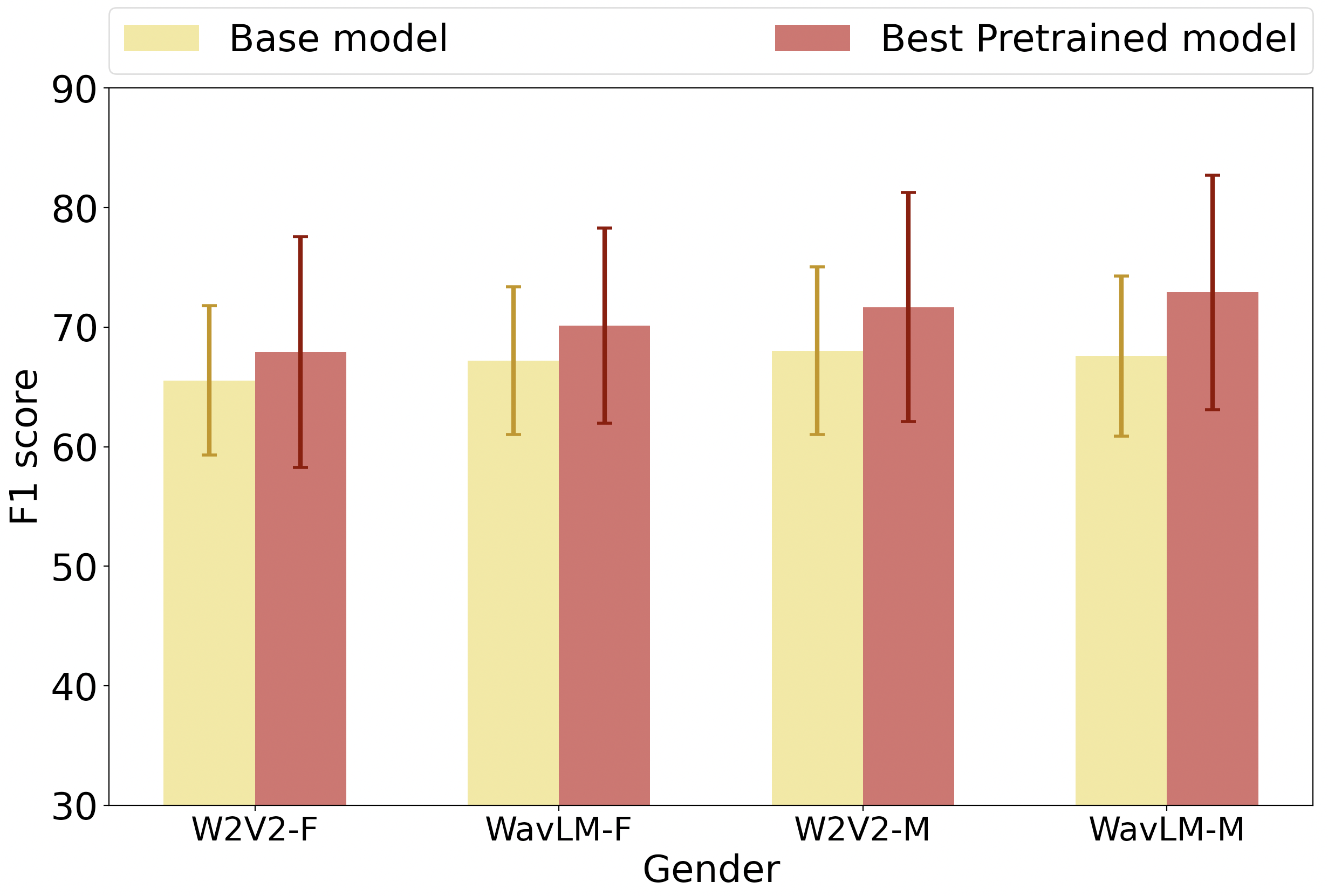}
    \vspace{-1.5ex}
    \caption{Gender based Child-adult classification F1 scores.}
    \label{fig:genderplot}
\vspace{-4.5ex}
\end{figure}

\subsection{Classification Evaluation}
\vspace{-1ex}

In this subsection, we analyze the experimental results reported in Table~\ref{tab:W2V2results} and Table~\ref{tab:WavLMresults} to address the following questions:

\noindent \textbf{Does pre-training with more child speech improve the classification?} The results reveal that pre-training with additional child speech improves the child-adult classification F1 score over the base model. This underscores the models' ability to account for the heterogeneity that is inherent in children's speech. It can be observed that WavLM-based pre-trained models show better performance compared to W2V2 across all the experiments. Both the classifiers show comparable performance, with the RNN-based classifier yielding the best score in the majority of the experiments. Among the datasets, the experimental results reveal better F1 scores in ADOSMod3 compared to the Simons corpus. One possible reason might be related to the session recording length difference between the datasets. The average duration of sessions in Simons corpus is much higher than ADOSMod3, resulting in greater potential variability and heterogeneity, which may have degraded the F1 scores. 

\noindent \textbf{Does pre-training with more transformer layers improve the classification?} It is interesting to note that, while in W2V2-based pre-training, the classification F1 keeps improving by tuning more transformer layers, in the case of WavLM, the performance improvements reach the maximum with tuning fewer transformer layers. One possible explanation is that the WavLM model is trained with an objective function to capture speaker related information, helping the model to achieve the optimum performance with lesser training. However, in both the scenarios, the model performance starts to degrade by adding more than four transformer layers. As these models are designed to provide generalized speech representations, tuning larger portions of these pre-trained models on a relatively smaller dataset might lead to the loss of generalizability, causing the performance to decrease for the classification task. However, our results provide compelling evidence that it is beneficial to adapt the last few transformer layers for the adult-child classification.

\noindent \textbf{Qualitative analysis} We present t-SNE visualizations of pre-trained embeddings for $2$ output classes from two sessions in Figure \ref{fig:tsne}. We plot the embeddings with and without additional pre-training. In both cases, it is evident from the plots that our method increases the discriminative information between them.

\subsection{Result Evaluation based on Demographics}
\vspace{-1ex}
\par For the gender-focused experiments, the relative improvement in F1 scores are $6.39\%$ and $3.14\%$ for the male and female subsets. Possibly due to both inherent speech pattern differences and inherent data distribution biases (see Section~\ref{sec:exp_demographic}), the models yield higher F1 scores in the male population than the female population. 
For the age-focused experiment, the relative improvements of $9.42\%$, $8.23\%$, and $4.06\%$ are seen for the three age-groups (youngest to oldest). The results imply that it is intrinsically challenging to model children of AG1 and AG2 due to the developing vocal tract behaviors among these ages. As a consequence, adding more children speech in training data provides greater benefits to the model to capture more relevant information, resulting in more improvement in the younger age-group than the older ones.


\begin{figure}
    \captionsetup{justification=centering}
    \begin{center}
    \includegraphics[width=0.45\textwidth]{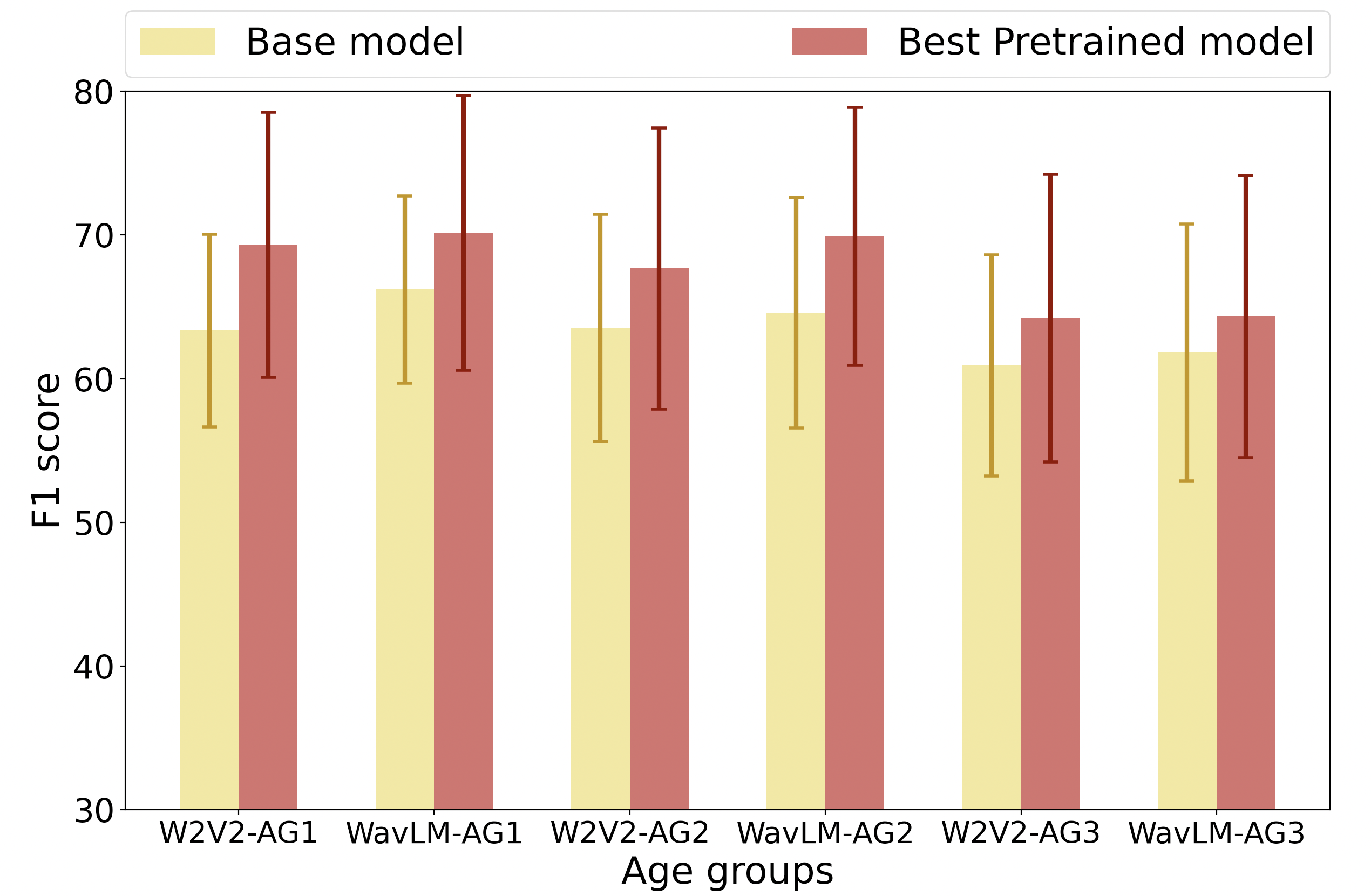}
    \vspace{-1.5ex}
    \caption{Age based Child-adult classification F1 scores.}
    \end{center}
    \label{fig:ageplot}
\vspace{-6.5ex}
\end{figure}

\vspace{-2.5ex}
\section{Conclusion}
\label{section:conclusion}
\vspace{-0.5ex}

Past work has demonstrated the promise of deploying self-supervised algorithms in a variety of downstream tasks like ASR, speaker diarization, and speaker verification~\cite{yang2021superb,fan2020exploring}. In this work, we investigate the utility of additional pre-training with more child speech, even in the presence of the inherent heterogeneity and variability, to improve child-adult speaker classification in clinical recordings involving interactions with children with autism. The experimental results with the proposed models  support our hypothesis of benefiting from incorporating child speech based additional pre-training, across both age and gender dimensions of variability. 

\par In this work, we used the manually-annotated ground truth labels for identifying and evaluating the speech and non-speech regions. In the future, we plan to build a child-adult diarization framework with an integrated \textit{Voice Activity Detection}~(VAD) system to further reduce the need of human effort.  
In addition, we plan to extend this study with an additional emphasis on early vocalization and speech  (from toddlers and infants) in the interaction. Unlike verbally fluent children, toddler speech contains significant amounts of pre-verbal sounds and non-verbal vocalizations, which pose additional challenges for automated processing.

\section{Acknowledgements}
\vspace{-1.5ex}
This work is supported by funds from USC Hearing, Communication and Neuroscience~(HCN) pre-doctoral fellowship, NIH and Simons foundation.
\bibliographystyle{IEEEtran}
\bibliography{main}

\end{document}